\documentclass[runningheads,a4paper]{llncs}
\usepackage{amsmath}
\usepackage{algorithm}
\usepackage[noend]{algpseudocode}

\usepackage{apacite}

\usepackage{amssymb}
\usepackage{graphicx}

\usepackage{url}
\urldef{\mailsa}\path|pmaguire@cs.nuim.ie|    
\newcommand{\keywords}[1]{\par\addvspace\baselineskip
\noindent\keywordname\enspace\ignorespaces#1}

\begin{document}

\mainmatter  


\title{Combining Independent Smart Beta Strategies for Portfolio Optimization}

\titlerunning{Combining Independent Smart Beta Strategies for Portfolio Optimization}

%
%
\author{Phil Maguire\textsuperscript{a,}%
\thanks{Corresponding author: pmaguire@cs.nuim.ie}%
\and Karl Moffett\textsuperscript{a}\and Rebecca Maguire\textsuperscript{b}}
%



\institute{\textsuperscript{a}Department of Computer Science, National University of Ireland\\
Maynooth, Ireland\\
\textsuperscript{b}Department of Psychology, National University of Ireland\\
Maynooth, Ireland\\
}

%
%

\toctitle{Lecture Notes in Computer Science}
\tocauthor{Authors' Instructions}
\maketitle

\begin{abstract}

Smart beta, also known as strategic beta or factor investing, is the idea of selecting an investment portfolio in a simple rule-based manner that systematically captures market inefficiencies, thereby enhancing risk-adjusted returns above capitalization-weighted benchmarks. We explore the idea of applying a smart strategy in reverse, yielding a ``bad beta'' portfolio which can be shorted, thus allowing long and short positions on independent smart beta strategies to generate beta neutral returns. In this article we detail the construction of a monthly reweighted portfolio involving two independent smart beta strategies; the first component is a long-short beta-neutral strategy derived from running an adaptive boosting classifier on a suite of momentum indicators. The second component is a minimized volatility portfolio which exploits the observation that low-volatility stocks tend to yield higher risk-adjusted returns than high-volatility stocks. Working off a market benchmark Sharpe Ratio of 0.42, we find that the market neutral component achieves a ratio of 0.61, the low volatility approach achieves a ratio of 0.90, while the combined leveraged strategy achieves a ratio of 0.96. In six months of live trading, the combined strategy achieved a Sharpe Ratio of 1.35. These results reinforce the effectiveness of smart beta strategies, and demonstrate that combining multiple strategies simultaneously can yield better performance than that achieved by any single component in isolation.


\end{abstract}

\section{Introduction}

According to the Capital Asset Pricing Model (CAPM), stock returns should be a linear function of beta. In other words, returns should reflect how risky a stock is relative to the market. The model defines the expected returns of a portfolio as:

~

$E(R_i) = R_f + \beta_i(E(R_m) - R_f) + \alpha$

~

where $R_f$ is the risk-free rate of return, $\beta_i$ is a measure of the correlation between portfolio $i$'s returns and that of a benchmark, and $R_m$ is the expected returns of that benchmark asset (Perold, 2004). The elusive $\alpha$ term is the excess returns that are generated from selecting long positions which outperform the market, or short positions that underperform. In an efficient market (see Malkiel \& Fama, 1970), the expected value of the alpha coefficient is zero. 

The CAPM prediction, namely that returns should be a linear function of beta, does not match observations. Recent work on smart beta has suggested that superior risk-to-reward performance can be achieved by following a few simple rules (e.g. Amenc, Goltz \& Martellini, 2013; Kahn \& Lemmon, 2016). For example, over the last 50 years low-volatility portfolios across the world have offered a desirable combination of high average returns and small drawdowns, bucking the intuition that risk should be compensated with higher expected profits (e.g. Haugen \& Baker, 1991; Clarke, de Silva \& Thorley, 2006; Baker, Bradley \& Wurlger, 2011), an effect that has been termed ``the greatest anomaly in finance'' (Baker et al., 2011).

Over the past five years, asset managers, including BlackRock, Legg Mason and Amundi, have launched a rapidly increasing number of smart beta funds, which act as a halfway house between active and passive management. Assets under management in smart-beta strategies ballooned from \$103bn in 2008 to \$616bn in 2015 (Mooney, 2016). Investors continued to pile into smart-beta funds in the first quarter of 2017, driving a 2,000\% rise in new money allocated to such strategies, and leading to warnings that it could all go horribly wrong (Mooney, 2017). Smart beta funds now account for one seventh of the \$4 trillion invested in the global ETF market. 

The smart beta concept holds that the intuitions of active managers can be outperformed by selecting portfolios according to a few simple rules, thus eliminating the substantial fees which eat into profits. Research has suggested that smart-beta strategies deliver outperformance in the long term, both empirically and theoretically (Mooney, 2016; see Amenc, Goltz, Martellini \& Retkowsky, 2011).

Nevertheless, there remains some debate as to whether smart beta is real, insofar as it can be reliably and consistently exploited in practice (e.g. Li, Sullivan \& Garcia-Feij\'{o}o, 2014; Malkiel, 2014). Many factors which appear to be supportive of smart beta may be nothing more than temporary anomalies that have been discovered through data mining (Arnott, Beck, Kalesnik \& West, 2016; cf. Asness, 2016). These anomalies may be small enough to be eroded by transaction costs, or they may reflect some statistical quirk such as future bias.  Arnott et al. (2016) suggest that smart beta strategies are not sustainable. Due to the soaring popularity of such funds, prices are being pushed up in a way that inflates past performance, leading to a ``smart beta bubble''.

Our study seeks specifically to address the question of whether applying a combination of smart beta strategies together can deliver real profits beyond the market benchmark. Accordingly, we apply a monthly reweighting strategy, which has minimal associated trading costs, as little rebalancing is required. We also apply our strategy out-of-sample, developing the portfolio based on historical data and applying it to real live data, thus removing any possibility of future bias. Finally, we investigate whether the combination of multiple independent strategies can enhance performance. If smart beta really does deliver superior returns, then combining a diversified set of smart beta strategies should deliver a risk to reward performance above any single strategy in isolation.

\section{Quantopian}

Our algorithms are tested via the online crowd-sourced hedge fund Quantopian. This service allows users to develop trading algorithms through an online Python-based research environment and a separate integrated development environment. Users are provided with years of minute by minute US stock pricing data, as well as fundamental business data and a full backtesting suite that can be used to test the past performance of an algorithm. The backtesting suite is designed to give an authentic representation of the performance of a trading system, featuring commission and slippage models that accurately account for the transaction costs involved in opening simulated positions, as well as the associated bid/ask spread costs. Monthly competitions are run which permit Quantopian members to pit their algorithms against those of other users. As of July 2016, Quantopian had about 85,000 members, with \$250 million under direction (Bradley, 2016).

\subsection{Characteristics of a Lucrative Hedge-Fund}

Quantopian grants capital allocations to the top performing algorithms which meet a set of strict requirements over a sustained trading period. The following desiderata are identified:

~

\textbf{Low Exposure to the Market:} Beta is a common metric used to calculate a portfolio's exposure to overall market movement. It measures the volatility of a portfolio and gives a feel of the systematic risk associated with an asset, describing the correlation between the portfolio's return and that of a benchmark asset. Price movements of an asset with a high beta coefficient are expected to reflect those of the benchmark asset.

~

$\beta =  (Covariance(R_i,R_m))/(Variance(R_m))$

~

Algorithms should have a beta coefficient of between -0.3 and +0.3 when compared to the returns of the S\&P 500. This can be achieved through appropriate risk management, and ensuring that the portfolio is hedged at all times.

~

\textbf{Consistent Profitability:} Another requirement is that portfolios generated by a trading algorithm should consistently demonstrate a Sharpe Ratio greater than 1. The Sharpe Ratio is calculated by subtracting the risk-free rate from a portfolio's returns and dividing it by the standard deviation of its return series. In other words, it is a measurement of expected return per unit risk of a portfolio. A high Sharpe Ratio indicates that a portfolio's returns are steady and generated with relatively little risk.

~

\textbf{Actively Trading Algorithms:} Portfolios are required by Quantopian to rebalance their capital at least once per month or at most twice a day. This constraint helps to ensure that portfolio performance reflects the quality of the algorithm, as opposed to just a lucky selection of stocks.

~

\textbf{Low Correlation to Peers:} The positions that an algorithm opens must have an average pairwise correlation of between -30\% and +30\% to those opened by other users' trading algorithms. In other words, a successful algorithm must contribute unique information.

\section{Beta-neutral portfolio}

In its truest form, a beta-neutral strategy is one that remains `market neutral' at all times, with its equity being evenly divided between long and short positions. Sorensen, Hua and Quin (2007) state that the market neutrality of a long-short equity portfolio leads to a higher risk adjusted return, making these portfolios more enticing to investors. 

If a portfolio is designed to be unaffected by overall directional market movement, having a market beta value very close to 0, then the beta term from the CAPM expected returns formula can be essentially eliminated. This leaves the alpha term as the only factor which can influence the returns, giving the alternative name `pure alpha' to these strategies. Returns are generated from picking out long and short positions from a universe of stocks which outperform and underperform the market. When forming a pure alpha trading approach, risk must be managed meticulously by hedging the long and short positions against each other.

Granizo-Mackenzie (2016) outlines an interesting idea for developing a long-short equity strategy. This involves taking a universe of stocks (e.g. the constituents of the S\&P 500), and ranking them by a predetermined factor. The trader then opens long positions on the top $N$ ranked stocks and short positions on the bottom $N$ stocks. Returns are generated from the divergence of the two groups of stocks. If the long positions collectively outperform the short positions, positive returns are generated; likewise if they underperform the short positions, negative returns are generated. One caveat is that it is imperative to choose a large enough value for $N$, so as to limit exposure to idiosyncratic risk. If a large value for $N$ is selected, then the cumulative beta of these positions will tend to 1, given that the portfolio reflects a substantial subset of the market. This is the smart beta portfolio. Likewise, the short positions, or ``bad-beta portfolio'', will have a beta close to -1 (also known as `alternative beta'; see Jaeger \& Pease, 2008). Combining these two portfolios together should therefore yield an overall beta value of close to 0.

Our idea is to use a novel smart beta strategy to carry out the ranking outlined by Granizo-MacKenzie (2016), namely momentum oscillators. In technical analysis, oscillators are tools used for checking trend reversals, and can be used both on price and volume. In brief, momentum oscillators constitute a set of calculations based on recent price trends that vary between a defined range to suggest where prices will move. For example, a recent uptrend suggests an upward movement of price over a certain period, while a downtrend suggests the opposite. Jawade, Naidu and Agrawal (2015) investigated the performance of a number of different momentum oscillators. They found that the Relative Strength Index and the Stochastic Oscillator were the most indicative oscillators in analysing trends. In contrast, Money Flow Index did not emerge as a viable oscillator in either uptrends or downtrends. They also highlighted the importance of analysing the volume of stocks traded during downtrends.

We also included another momentum indicator known as Moving Average Crossover, which takes the average closing price of an asset over two different windows, and returns a ratio of the shorter period average with respect to the longer period (Granizo-MacKenzie, 2016). The four indicators used to predict the current market trend, namely Relative Strength Index, Stochastic Oscillator, Moving Average Crossover and Volume are described in more detail below.

~

\textbf{Relative Strength Index (RSI):} Relative Strength Index is an indicator first developed by Wilder (1978). The index aims to measure the rate at which a stock's price is rising or falling by analysing the gains and losses over a certain lookback window. It assigns a value of between 0 and 100 to the stock's latest behaviour and is formulated as follows:

~

$RSI = 100 - 100/(1+RS)$

~

with Relative Strength (RS) being defined as: \textit{RS = (Average of x day's gains)/( average of x day's losses)}. In our analyses we calculated relative strength over a 14-day window.

The RSI indicator is one of the most popular trading tools used by technical analysts. While these analysts focus on the graphical aspect, the RSI values themselves can be used to identify trends (Rudik, 2013).

~

\textbf{Stochastic Oscillator:} The Stochastic Oscillator consists of two different signals known as $\%k$ and $\%D$ signals. The $\%D$ is simply a 3-period moving average of the $\%k$ signal which helps to `smooth' it out.

~

\textit{\%k = (Latest Closing Price - Lowest Price of last N days)/(Highest price of last N days - Lowest price of last N days)}

~

$N$ can be typically assigned a range of different values depending on the trading style. We adopted a 21-day time frame (the average number of days in a trading month) to calculate the $\%k$ signal, thus mirroring the time period over which we would be analysing stock returns. We opted to use only the $\%D$ signal generated by the Stochastic Oscillator, as it would be less sensitive to noise, and thus better suited to the machine learning process.

Like the RSI indicator, the $\%k$ and $\%D$ signals generate a value between 0 and 100. The stochastic oscillator signals are traditionally used by charting them alongside a time series of a stock's price, with `buy' and `sell' signals being generated when certain criteria are met (see Ni, Liao and Huang, 2015).

~

\textbf{Moving Average Crossover:} Moving Average Crossover is among the most popular technical analysis tools used by traders (Cai et al., 2010). The idea is closely related to that of the Moving Average Convergence Divergence (MACD) technical analysis tool. Though it is more traditionally used by plotting a chart of MACD values alongside a time series of a stock's price, we decided to analyse the latest value of the MACD. Granizo-Mackenzie (2016) identifies this value as a measure of momentum, but warns that choosing a specific set of values when analysing charts can cause overfitting. Keeping this measure of momentum as general as possible, we opted to use a single month moving average divided by a two-month moving average as a measure of momentum.

~

\textit{MA$_{crossover}$ = (1-month simple moving average)/(2-month simple moving average)}

~

\textbf{Volume:} As opposed to analysing the raw volume of each stock traded, Quantopian provides a built-in factor, known as the average dollar volume factor, which provides a simple method of standardizing volume data by disregarding the individual price of a stock. It takes the average volume that a stock has traded over a lookback window and multiplies it by the stock's latest share price. According to Quantopian, this is a better measure of a stock's liquidity than raw volume alone. Remorov (2014) examined the influence of trading volumes on stock prices during market crashes, finding a relationship between volumes and price during these downturns in the market.

~

The simplest way of normalising the data was to take the rank of each feature with respect to all the other stocks (i.e. giving 1 to the stock with the highest value, and 500 to the lowest).

\subsection{Model Selection}

Using the machine learning scikit-learn library (see Pedregosa et al., 2011), we instantiated a range of different classifiers in the research environment and performed cross validation with each classifier to assess the predictive robustness (see Browne, 2000).

The same training and test sets were used for each classifier, though we used a number of different training and test sets from different periods, performing cross validation multiple times to get a true reflection of each classifier's performance. We found that the AdaBoostClassifier was the most effective of these classifiers (see R{\"a}tsch, Onoda \& M{\"u}ller, 2001). AdaBoost, short for Adaptive Boosting, is a machine learning meta-algorithm originally developed by Freund and Schapire (1999), which can be less susceptible to overfitting than other algorithms. 

With a cross validation accuracy of 53\%, the margin of success was fine, but if used wisely, could prove significant. We found that training the classifier with data from the previous two months was the optimal training period. Using a longer period than two months meant that the classifier was too slow to respond to current market trends, as well as spanning multiple market trends, resulting in ineffective learning. On the contrary, using only a single month of training data failed to provide enough samples to learn adequately.

\subsection{Fundamental Factors}

Stone, Chen and White (2014) investigated the influence that various factors have on the performance of a stock, and outlined 8 key categories to be considered: valuation, growth, quality, efficiency, momentum, risk/size, pay-out and profitability. In their paper, they define over 50 financial factors which can be used, and divide them by category.

Quantopian provide fundamental business data, as supplied by Morningstar. We added all of the factors to the training sets which were possible to implement in the live trading environment. We then retrained the classifier with the new data and analysed the importance of each added feature. The initial momentum factors made up the top 3 features, showing that they are the most consistent predictors. Although the Average Dollar Volume factor scored lower, this is likely because the period during which the factors were tested was a time when the market was experiencing an uptrend: volume is not particularly useful during uptrends but critical when analysing performance during downtrends (Remorov, 2014). 

We developed a version of the machine learning algorithm that included some fundamental factors, along with momentum and volume indicators. Including the factors which have been deemed to be most important (see Hsu \& Kalesnik, 2014) did not enhance performance. Based on these inconsistent results, we decided that it was not worthwhile including any fundamental factors. In line with our observations, Jawade, Naidu, and Agrawal (2015) have argued that fundamental factors cannot be used to anticipate stock price movement, since these factors have already influenced the current stock price.

\subsection{The Effects of Unstable Trading Periods}

Throughout both research and development of the model, we noticed that the algorithm performed poorly immediately after extremely unstable trading periods. Because its performance \textit{during} these periods wasn't noticeably different than any other, we deduced that the poor performance was a result of poor learning from these highly volatile datasets.

Analyzing monthly high, low and close values of the S\&P 500 from 1997 - 2007, we examined the log normalized monthly ranges (i.e. max - min / average) to make inferences about the effects of unstable periods (see Figure 1). 

\begin{figure}[H]
\centering
\hspace*{-1cm} 
\includegraphics[trim = 10mm 80mm 5mm 5mm, clip, width=14cm]{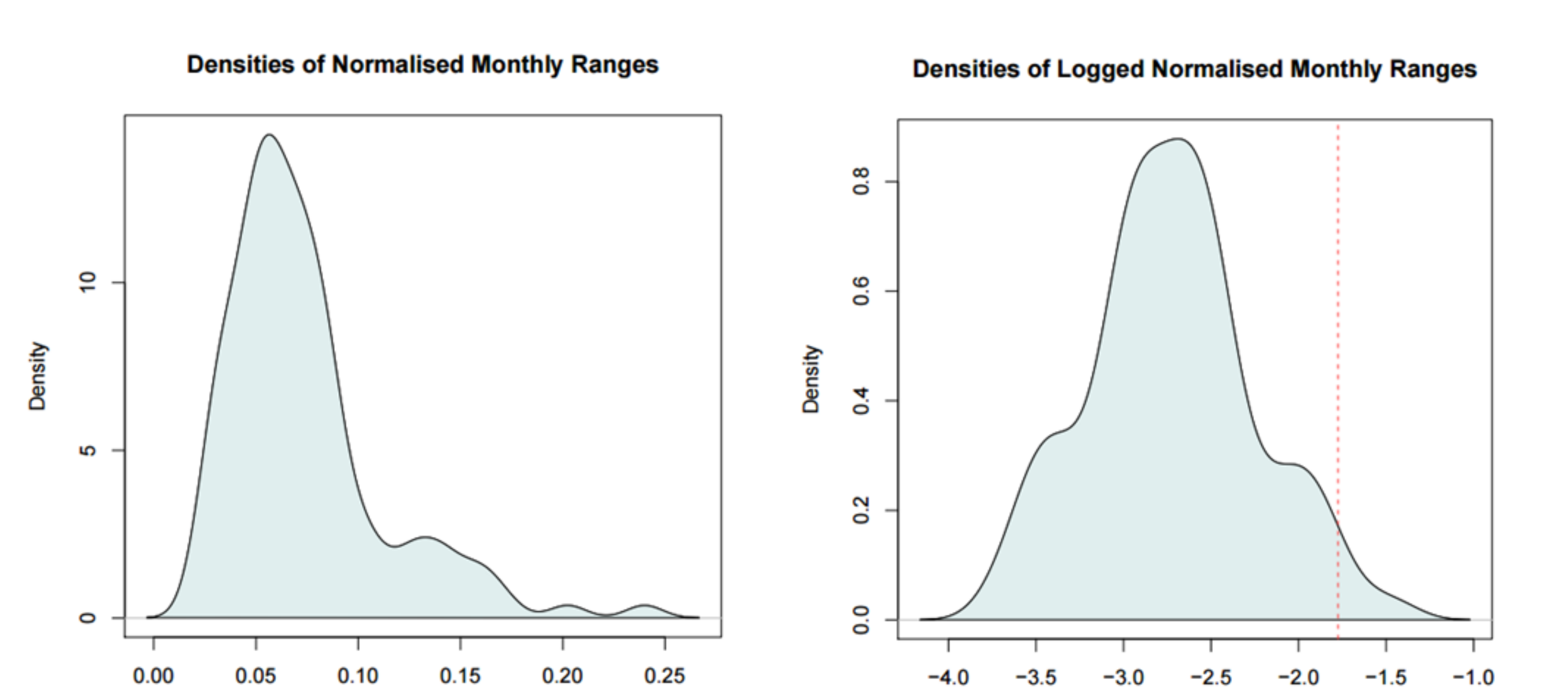}
\caption{Distribution of S\&P 500 monthly ranges and natural log of these values}
\label{fig:example}

\end{figure}

We found that omitting the most unstable trading periods (i.e. those with a range greater than 17\%, as indicated by the dotted line in Figure 1) from the learning process resulted in an increase in performance. Such high volatility periods are transient, do not persist, and hence are not representative of the current market trend.

\subsection{Performance of Momentum Oscillator Quintiles}
By taking our universe of stocks, namely the 500 constituents of the S\&P 500, and creating a portfolio of evenly weighted long positions for each of the 5 quintiles, we could evaluate the performance of the ranking ability of our machine learning classifier. Each quintile was rebalanced monthly, following the latest data and associated predictions. 

In Figure 2 below, we can clearly see a relationship between quintiles and performance. This is evidence that the use of momentum oscillators provides a valid stock ranking system which can be applied as part of a long-short equity strategy.

\begin{figure}[H]
\centering
\hspace*{-1cm} 
\includegraphics[height=10cm]{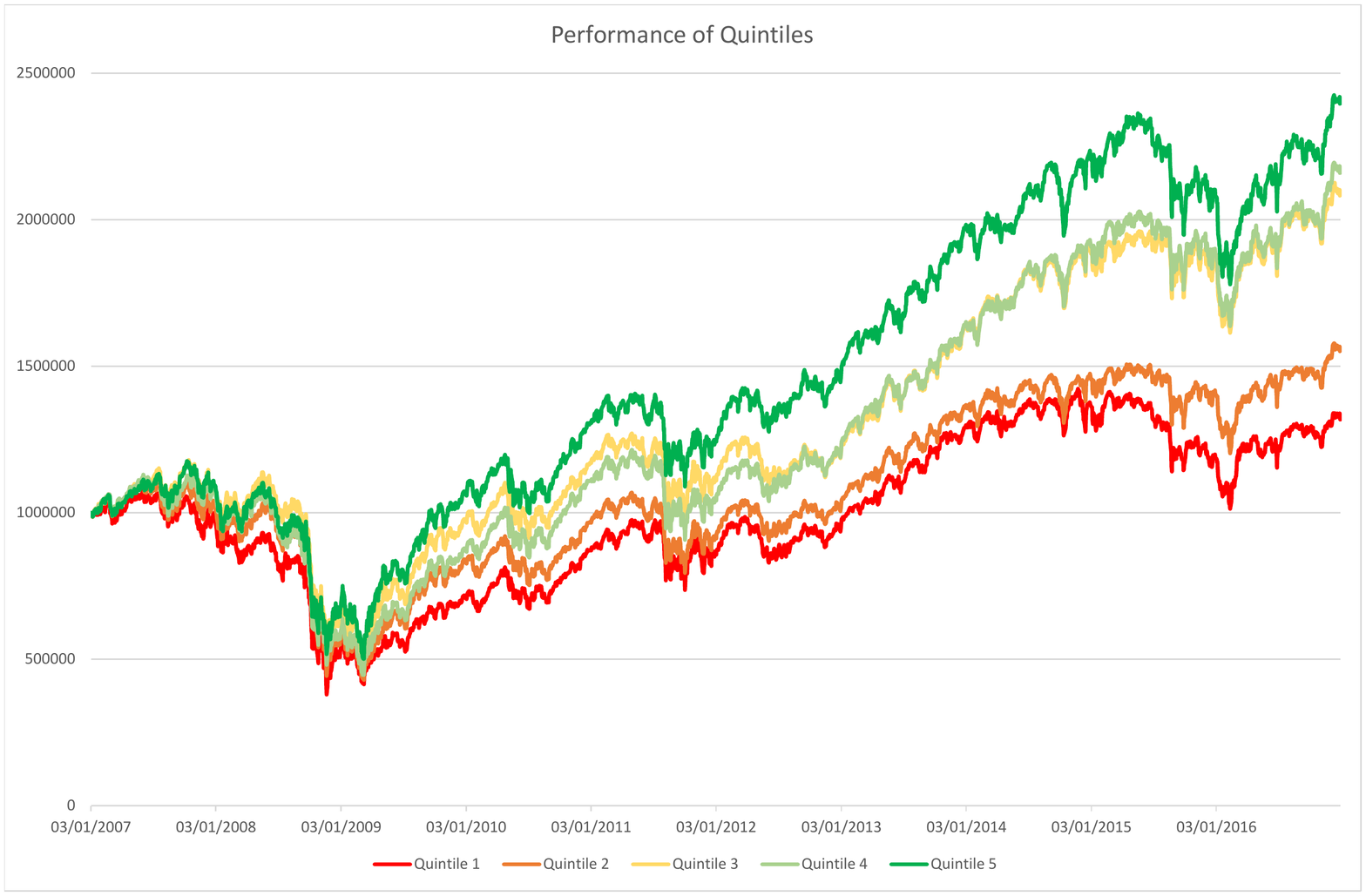}
\caption{Performance of each quintile as predicted by the machine learning classifier}
\label{fig:example}
\end{figure}

\section{Minimum Variance Portfolio}

Of all smart beta strategies, low-volatility investing has particularly strong empirical support. Based on a broad sample of international developed markets, Ang et al. (2009) found that stocks with recent past high idiosyncratic volatility had lower future average returns. Across 23 markets, the adjusted difference in average returns between highest and lowest quintile portfolios, sorted by volatility, was -1.31\% per month. This effect was found to be individually significant for every G7 country (Canada, France, Germany, Italy, Japan, the United States, and the United Kingdom), suggesting that the relation between high idiosyncratic volatility and low returns is not just a sample-specific or country-specific effect, but a global phenomenon.

Blitz and van Vliet (2007) provided further empirical evidence that stocks with low-volatility earn higher risk-adjusted returns than the market portfolio, even after controlling for well-known effects such as value and size. They found that the annual alpha spread of global low versus high-volatility decile portfolios amounted to 12\% over the 1986-2006 period, observing independent effects in the US, European and Japanese markets. 

Baker and Haugen (2012) analysed 33 different markets during the time period from 1990-2011, including non-survivors. They computed the volatility of total return for each company in each country over the previous 24 months, ranking stocks by volatility and grouping them into deciles. In each one of the 21 developed countries, the lowest volatility decile had both lower risk \emph{and} higher return, leading to substantial divergence in Sharpe Ratios. 

In another study, Clarke, de Silva and Thorley (2006) found that minimum variance portfolios, based on the 1,000 largest US stocks over the 1968-2005 period, achieved a volatility reduction of about 25\%, while delivering comparable, or even higher, average returns than the market portfolio.

The concept of volatility is closely related to that of beta: beta describes the volatility of an asset relative to the market, in essence the correlated relative volatility (by definition the market has a beta of 1). In line with other results, Black (1993) found that, in the period from 1931 to 1965, low-beta stocks in the U.S. did better than CAPM predicts, while high-beta stocks did worse.

Frazzini and Pedersen (2014) provided empirical evidence that portfolios of high-beta assets have lower alphas and Sharpe Ratios than portfolios of low-beta assets. They found that high beta is associated with low alpha for US equities, 20 international equity markets, treasury bonds, corporate bonds, and futures. They also found that a betting-against-beta (BAB) factor, which is long leveraged low-beta assets and short high-beta assets, produces significant positive risk-adjusted returns, and rivals standard asset pricing factors (e.g., value, momentum, and size) in terms of economic magnitude, statistical significance, and robustness across time periods, subsamples of stocks, and global asset classes.

Frazzini and Pedersen (2014) hypothesized that constrained investors stretch for return by increasing their betas, thus artificially enhancing the price and lowering the value of high beta securities. This hypothesis is supported by the observation that both mutual funds and individual investors tend to hold securities with betas that are significantly above one (Frazzini \& Pedersen, 2014). In contrast, leveraged buyout funds and Berkshire Hathaway, all of which have access to leverage, tend to buy stocks with betas below one. These investors take advantage of the BAB effect by applying leverage to safe assets and are compensated by investors facing borrowing constraints who take the other side. According to Frazzini and Pedersen (2014), Warren Buffett gets rich by betting against beta, that is, buying stocks with betas significantly below one and applying leverage.

In light of this substantial evidence, we decided to employ a second smart beta strategy, based on minimizing volatility.

\subsection{Minimum Variance Portfolio Algorithm}

Mostowfi and Stier (2013) state that many portfolio managers have turned to Minimum Variance Portfolios (MVPs) due to their straightforward calculation using a covariance matrix of historical asset returns. Based on Frazzini and Pedersen's (2014) work, we opted to use a selection of low beta stocks in our MVP. Beta values were calculated with respect to the returns of our default benchmark asset, the S\&P 500 index, using a lookback window of 66 days, which represents about 3 months of trading activity. We selected a relatively small number of these low beta stocks, namely the lowest 25, to comprise our MVP. Herssein's (2016) procedure was used for deriving the appropriate allocations of each asset, using only historical returns. 

To begin, a matrix of the daily historical returns of each asset from the previous quarter of trading is required. The covariance matrix of this returns matrix is then calculated ($V$). Then by simply using a column vector of ones ($I$), and a column vector of the average returns of each asset ($R$), we can calculate the weights of the Minimum Variance Portfolio as follows:

~

$mvpweights = V^{-1}R + V^{-1}I$

~

where $V^{-1}$ is the inverse of the calculated covariance matrix. We decided not to place any constraints on the MVP, such as restricting the maximum allowable allocation. The nature of this style of MVP allows both long and short positions, the valence of which is ultimately decided by whether the historical returns are positive or negative.

The average exposures are 65\% long, 35\% short, given that the market tends to rise. However, it is noticeable over periods where the market is in decline that the exposures are closer to 55\% long, 45\% short. 

\section{Results and Evaluation}

We opted to use a 10-year trading period from the beginning of 2007 up to the end of 2016 as the backtesting period. Testing over a sustained period of time gives a clear indication of the validity of a trading strategy, with this decade capturing a variety of both uptrends and downtrends in the market. The S\&P 500 index was used as the benchmark asset to compare the performance of each strategy.

\subsection{Momentum Oscillator Performance}

First we tested the long-short beta-neutral momentum oscillator strategy. We opened evenly-weighted long positions on the top ranked quintile and evenly weighted short positions on the bottom ranked quintile. Designating half of capital to long positions and the remaining 50\% to short positions ensured that this portfolio was dollar neutral, meaning that the resulting portfolio beta value was very close to 0. Although the portfolio showed rather paltry annual returns of only 2.5\%, inspecting the nature of the returns allowed us to draw more meaningful conclusions about the strategy. 

The portfolio's negligible beta value of -0.01 demonstrates that this is truly a `pure alpha' strategy. All returns are completely independent of overall market movement and are instead
 generated through intelligent stock picking. The portfolio's maximum drawdown (i.e. largest loss) of 8.7\% shows how low-risk this portfolio is, compared to the S\&P 500 with a max drawdown of over 50\% (see Figure 3). The performance was largely unaffected by the market crash which occurred between late 2007 and early 2009.

Stock markets globally are generally expected to rise into the future, this being perhaps their only genuinely predictable feature (Maguire et al., 2017). This fact could be used to reap further returns from our algorithm by allocating a larger proportion of capital into long positions than into short positions. Although it leaves the portfolio slightly exposed to directional market movement, the hedging can be carried out in a controlled manner so as to stay within a target beta value.

\begin{figure}[H]
\centering
\hspace*{-1cm} 
\includegraphics[height=10cm]{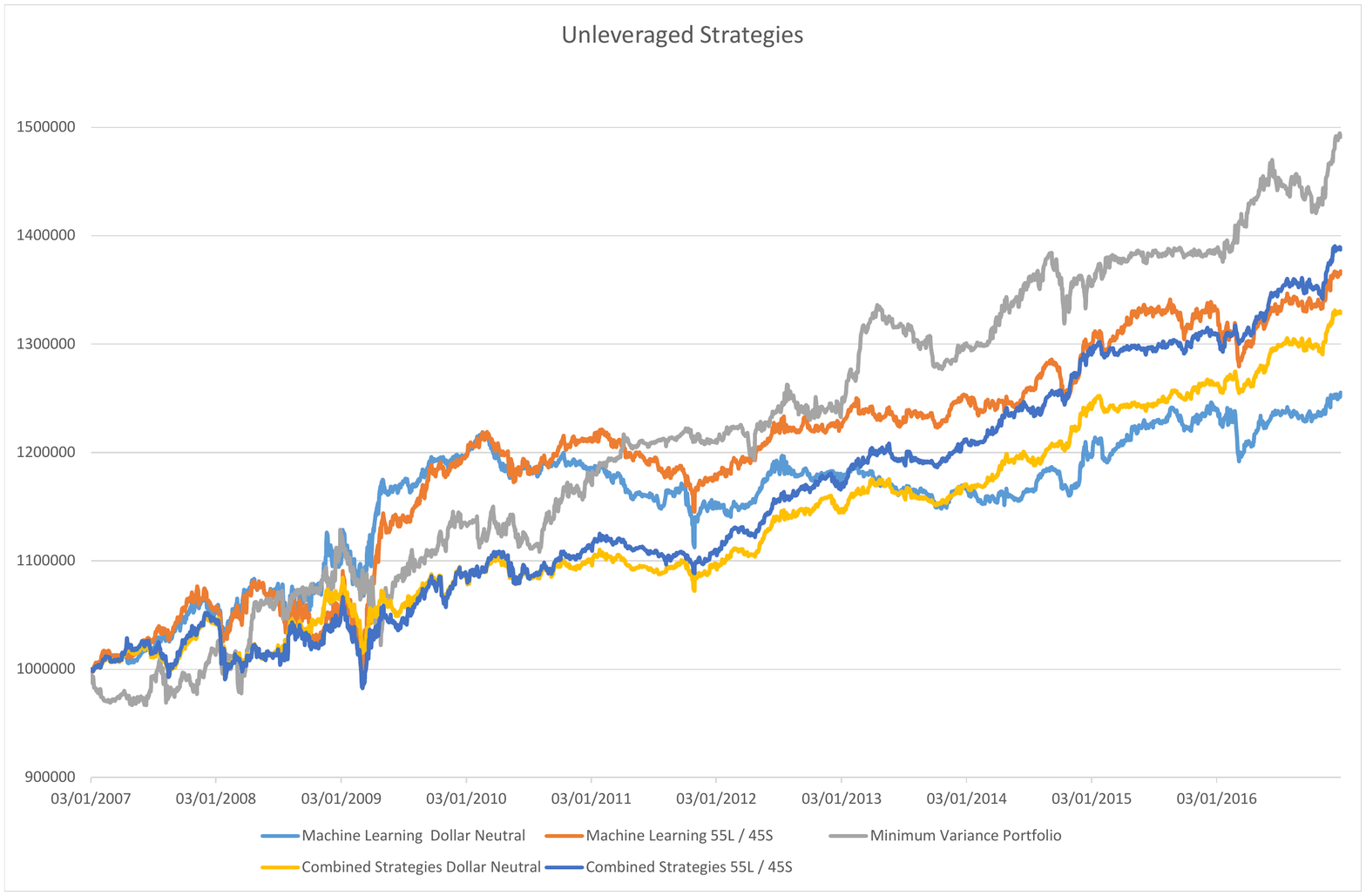}
\caption{Performance of each unleveraged strategy}
\label{fig:example}
\end{figure}

Assigning 55\% of our capital to long positions and the remaining 45\% into short positions, performance is substantially enhanced (see Figure 3). Overall returns across the 10-year testing period are boosted by over 10\%, and the Sharpe Ratio increases from 0.58 to 0.70. The strategy retains a low market beta coefficient of 0.1, which is comfortably within the allowable bounds for Quantopian. Surprisingly, the maximum drawdown of this portfolio is only 8.5\%, as opposed to the dollar neutral portfolio's corresponding value of 8.7\%. 

\begin{table}

  \begin{center}
    \caption{Profitability and risk metrics of unleveraged strategies compared to benchmark}
\begin{tabular}{| l | c | c | c | c | c | }
\hline			
  \bf{} & \bf{Returns (\%)} & \bf{Sharpe Ratio} & \bf{Beta} & \bf{Max
Draw (\%)} & \bf{Vol. (\%)}\\
  \hline
S\&P 500 & 90.2 & 0.42 & 1 & 54.9 & 20\\
   \hline
$\beta$-Neutral & 25.4 & 0.58 & -0.01 & 8.7 & 4\\
   \hline
$\beta$-Neutral
(55\%) & 36.5 & 0.70 & 0.1 & 8.5 & 4\\
   \hline
MinVar & 54.1 & 0.89 & 0.06 & 8.9 & 5\\
   \hline
Combo & 33 & 0.9 & 0.03 & 6.4 & 3\\
   \hline
Combo (55\%) & 38.8 & 0.92 & 0.08 & 7.9 & 4\\
   \hline          
  \end{tabular}
  \end{center}
\end{table}

\subsection{Minimum Variance Portfolio Performance}

The Minimum Variance Portfolio greatly outperformed the momentum oscillator algorithm in terms of returns and Sharpe Ratio, while only being marginally more volatile. The decade trading period yields a 54.1\% return of investment as well as a healthy Sharpe Ratio of 0.89. The portfolio's market beta of only 0.06 reveals a miniscule exposure to the benchmark asset, while experiencing a max drawdown of only 8.9\% over the testing period. The beta value here is calculated as an average value over the 10 year period, taking the daily returns of the portfolio and comparing it with the daily returns of the S\&P 500. With the MVP being dynamically weighted, and given the low beta coefficient, we can deduce that the vast majority of returns are generated from a combination of both clever stock picking and intelligent asset weighting.

\subsection{Combined Strategies Performance}

Combining the two strategies and running them in parallel enhanced performance even further. Although absolute levels of return were necessarily lower than those of the MVP in isolation (up to 38.8\% for the strategy with 55\% assigned to long positions; see Figure 3), the level of volatility was lower, making this product more suitable for leverage. This performance is reflected in a Sharpe Ratio of 0.92.

\subsection{Leveraged Results}

Leveraging is the idea of borrowing money with the intention of immediately re-investing this capital into a trading strategy. This multiplies the capital base available to a trader, effectively magnifying any gains and losses. Leveraging is best suited to strategies that are truly low risk, in which a trader has great confidence. Even if an algorithm only generates small returns, as long as they are consistent and produced through low risk techniques, then the strategy can be suitably leveraged to improve the absolute returns. Trying to find a balance between magnifying the absolute returns and increasing the max drawdown and other risk metrics, we decided on a leveraging factor of 2.

\begin{table}

  \begin{center}
    \caption{Profitability and risk metrics of leveraged strategies (by a factor of 2:1)}
\begin{tabular}{| l | c | c | c | c | c | }
\hline			
  \bf{} & \bf{Returns (\%)} & \bf{Sharpe Ratio} & \bf{Beta} & \bf{Max
Draw (\%)} & \bf{Vol. (\%)}\\
  \hline
S\&P 500 & 90.2 & 0.42 & 1 & 54.9 & 20\\
   \hline
$\beta$-Neutral & 58.7 & 0.61 & -0.03 & 16.8 & 8\\
   \hline
$\beta$-Neutral
(55\%) & 86.27 & 0.73 & 0.19 & 16.5 & 9\\
   \hline
MinVar & 131.4 & 0.9 & 0.13 & 17.2 & 10\\
   \hline
Combo & 80.1 & 0.94 & 0.05 & 12.6 & 6\\
   \hline
Combo (55\%) & 95.5 & 0.96 & 0.16 & 15.5 & 7\\
   \hline          
  \end{tabular}
  \end{center}
\end{table}

\begin{figure}[H]
\centering
\hspace*{-1cm} 
\includegraphics[height=10cm]{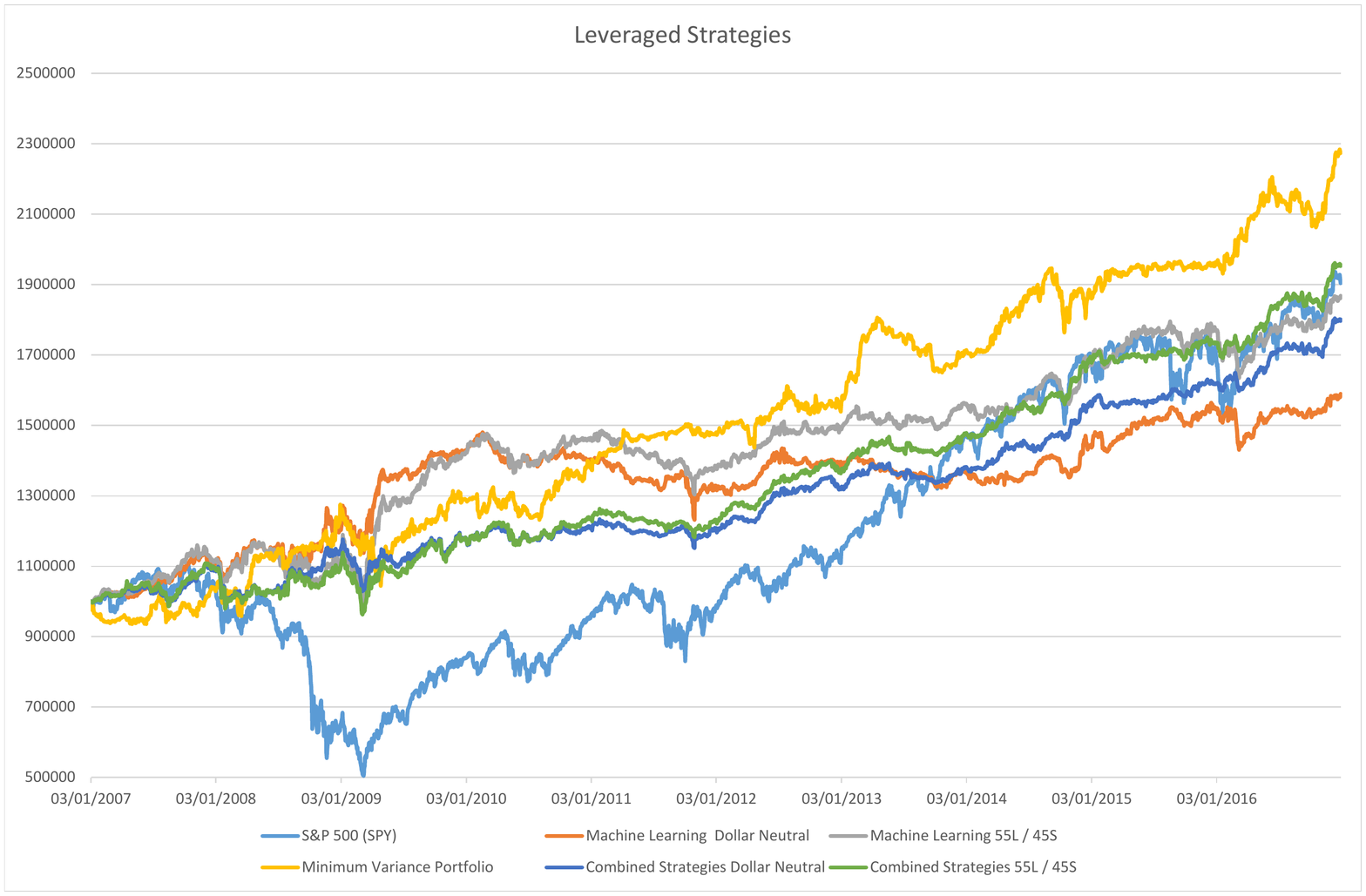}
\caption{Performance of each leveraged strategy and benchmark (S\&P 500)}
\label{fig:example}
\end{figure}

The use of leverage magnifies the net returns and beta coefficient by the leveraging factor, as well as the portfolio's max drawdown and annual volatility, while the Sharpe Ratio is relatively unaffected. The reason for a slight increase in the Sharpe Ratio for a leveraged investment is that the excess positive returns are themselves leveraged as they are earned, leading to a more rapid increase in profit, which offsets the costs of obtaining financing for leverage.

Figure 4 shows a wider range of returns from these leveraged portfolios. What is clear is that these smart beta strategies can outperform the S\&P 500. Table 2 reveals that the MVP has a lower Sharpe Ratio than both of the combined strategies, implying that the MVP returns have been generated by taking larger risks. For example, it experiences a sizably larger maximum drawdown, and has a persistently higher annual volatility. Combining independent smart beta strategies is desirable because it lowers risks, while maintaining profits.

\subsection{Live Trading and Competition Performance}

Although only a fraction of Quantopian's 100,000 users enter the monthly competitions, the standard of algorithm at the top of the leaderboards remains extremely high. Algorithms are evaluated mainly on their performance during a 6-month course of out-of-sample simulated live trading, and partially on a two year backtest prior to the start date. A strategy is scored on a range of profitability and risk metrics. Each metric is calculated as a rank among competitors, and an average of these scores is taken.

We entered our combined 55-45 momentum oscillator / MVP smart beta strategy into the Quantopian trading competition. Twenty-six weeks later, the algorithm stands at 84th position in the leaderboard, having outperformed thousands of other algorithms globally. Unleveraged, it currently boasts returns of 1.2\% since going live, with an excellent Sharpe Ratio of 1.35. 

Future work before capital allocation will include:

1. Ensuring that the portfolio stays dollar neutral at all times, which might involve replacing the MVP with a long-short betting-against-beta factor (see Frazzini \& Pedersen, 2014).

2. Ensuring that the portfolio stays sector neutral, with a maximum exposure for each sector of no more than a 60/40 long-short position. For example, the portfolio should not go long on all of its tech or energy stock components.

3. Shortening the rebalancing period to once per week, as opposed to monthly, in order to constrain any deviations from neutrality. This may incur further rebalancing costs.

\section{Conclusion}

In this study we have combined a long-short momentum oscillator strategy with a minimum volatility strategy. Evaluating the performance of the leveraged portfolios, we can see that the combined strategy outperforms the benchmark asset (S\&P 500) in terms of both net returns and other risk and profitability metrics. The strategy with the highest Sharpe Ratio achieves over 9\% annual returns, marginally more than the benchmark, but with significantly less risk. If we were willing to increase our risk tolerance, we could leverage this portfolio by a greater factor (e.g. 2.5:1) and substantially outperform the market. With returns of just under 140\% from 2007 to 2017, such a strategy would have outperformed the market by over almost 50\% (see Figure 5). The maximum drawdown is 19.1\%, considerably smaller than the market's corresponding value of over 50\%.

\begin{figure}[H]
\centering
\hspace*{-1cm} 
\includegraphics[height=10cm]{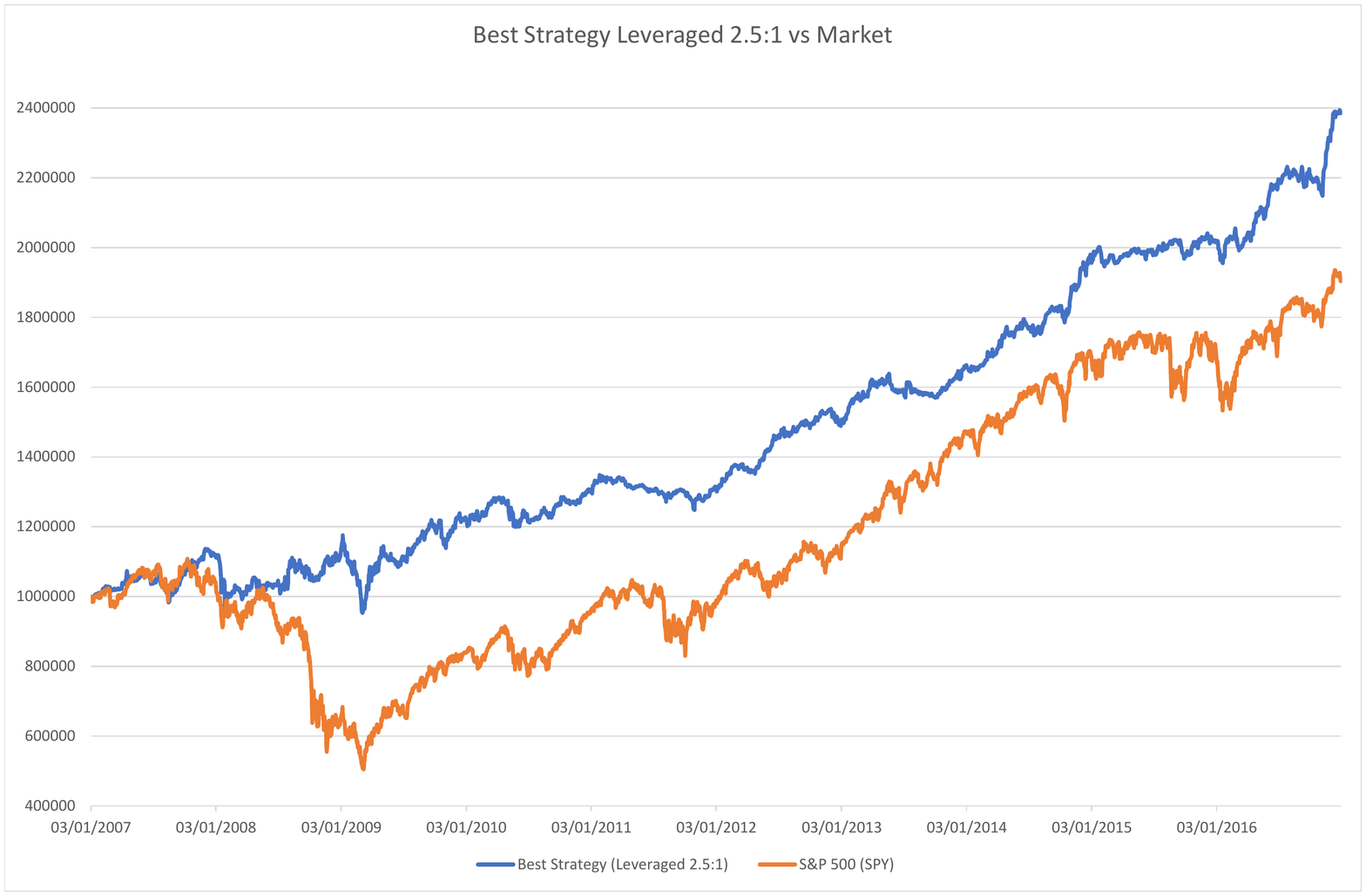}
\caption{Comparison of Best Strategy Leveraged vs S\&P 500}
\label{fig:example}
\end{figure}

While adhering to Quantopian's strict competition rules over our 10-year testing period, we can see that our strategy falls just short of achieving the recommended Sharpe Ratio of 1, managing 0.96. Nevertheless, it stays well within the beta bracket of +0.3 and -0.3, keeps leverage restrained below 3 and doesn't experience any excessively large drawdowns, all of which are desirable characteristics. 

In conclusion, the success of the algorithm in competitive live trading provides evidence that smart beta strategies can be effective, and that combining multiple smart beta strategies can be even more so. 

\nocite{*}

\bibliographystyle{apacite}

\setlength{\bibleftmargin}{.125in}
\setlength{\bibindent}{-\bibleftmargin}

\bibliography{ReferencesTOPICS}

\end{document}